\documentclass[twocolumn,showpacs,preprintnumbers,amsmath,amssymb]{revtex4}
\usepackage{graphicx}
\usepackage{dcolumn}
\usepackage{bm}

\begin{document}

\title{Decoherence of a Measure of Entanglement}

\author{Denis~Tolkunov}\email{tolkunov@clarkson.edu}
\author{Vladimir~Privman}\email{privman@clarkson.edu}
\affiliation{Center for Quantum Device Technology, Department of
Physics, Clarkson University, Potsdam, New York 13699--5721}
\author{P.~K.~Aravind}\email{paravind@wpi.edu}
\affiliation{Department of
Physics, Worcester Polytechnic Institute,
100 Institute Road, Worcester, MA 01609-2280}

%\date{\today}

\begin{abstract}

We demonstrate by an explicit model calculation that the decay of entanglement
of two two-state systems (two qubits) is governed by the product of the factors
that measure the degree of decoherence of each of the qubits, subject to independent
sources of quantum noise. This demonstrates an important physical property that separated
open quantum systems can evolve quantum mechanically on time scales larger than the times
for which they remain entangled.

\end{abstract}


\maketitle

Entanglement of quantum-mechanical states, referring to the
nonlocal quantum correlations between subsystems, is one of the
key resources in the field of quantum information science. Many
protocols in quantum communication and quantum computation are
based on entangled states [\onlinecite{Nielsen}]. When one
considers practical applications of entanglement, the coupling of
the quantum system and its subsystems to the environment,
resulting in decoherence, should be taken into account. It is
known [\onlinecite{Bennett, Peres}] that entanglement cannot be
restored by local operations and classical communications once it
has been lost, so understanding of the dynamics of decoherence of
entanglement is of importance in many applications.

There are two basic issues in the physics of the loss of
entanglement by decoherence, that, while intuitively suggestive,
thus far have allowed little quantitative, model-based
understanding. To define them, let us refer to two subsystems,
$S^{\rm (1)}$ and $S^{\rm (2)}$, of the combined system, $S$. The
first property of interest is the expectation that when the
systems are separated in that they are subject to independent
sources of noise, e.g., when they are spatially far apart, then
the decoherence of entanglement is faster [\onlinecite{Eberly,Eberly1,Eberly2}]
than the loss of coherence in the quantum-mechanical behavior of
each of the subsystems. Thus, the subsystems can for some time
still behave approximately in a coherent quantum-mechanical
manner, but without correlation with each other.

In order to define the second property of interest, let us point
out that the definition of ``decoherence'' of an open quantum
system is not unique. One has to consider the overall
time-dependent behavior of the reduced density matrix of the
system, obtained for a model of the environmental modes, which are
the source of noise and are traced over. This time dependence can
involve an oscillatory behavior corresponding to the initial
regime of approximately coherent evolution, with frequencies
determined by the energy gaps of the system (which can be shifted
by the noise). At the same time, there will be irreversible,
decay-type time dependencies manifest for larger time scales,
which can in many cases be identified with processes such as
relaxation, thermalization, pure decoherence, etc., that represent
irreversible noise-induced behaviors [\onlinecite{Louisell,
vanKampen, Blum, Abragam, Grabert, Leggett, Palma, Adiabatic,
Privman, Tolkunov}].

One, by no means unique, way to quantify the degree of loss of
coherence is by the decay of the absolute values of
off-diagonal elements of the reduced density matrix. This
definition is only meaningful at relatively late stages of the
dynamics, when the density matrix has already become nearly
diagonal in a basis favored by external and internal interactions, and
by environmental influences, e.g., for thermalization, the energy
basis. More careful definitions of measures of decoherence are
possible [\onlinecite{Fedichkin}], but we will use the
off-diagonal-element nomenclature for clarity. Recent experimental
NMR studies [\onlinecite{NMR}] have considered various ``orders of
coherence'' that involve off-diagonal elements, for systems of up
to 650 spins.

The second property of interest, is formulated in this language as
follows. For noninteracting and nonentangled subsystems, the
density matrix of the whole system will be a direct product of the
subsystem density matrixes. In this simple case, there will be
far-off-diagonal density matrix elements of the system that will
decay by a factor that is a product of the decay factors of the
subsystem off-diagonal elements. Specifically, if the large time
decay is exponential, then the decay {\it rates\/} will be
additive [\onlinecite{Wilhelm, Hilke}].

A related ``additivity'' property has been mathematically explored
for certain measures of initial decoherence
[\onlinecite{Fedichkin}], for {\it entangled\/} subsystems.
Recently, exploration of the following
physically very suggestive question has been
initiated [\onlinecite{Eberly1}]: If we know the suppression
factors, $0\leq \delta^{(1,2)}\leq 1$, that roughly measure
decoherence for the two subsystems, then are there any {\it
physically meaningful\/} quantities that are suppressed by the
product $\delta^{\rm (1)} \delta^{\rm (2)}$? The other suggestive
alternative is that the ``worst case scenario'' for physically
relevant loss-of-coherence measures of the combined system is
suppression by the factor of $\min ( \delta^{\rm (1)} ,
\delta^{\rm (2)} )$. The two alternatives are, of course, only
approximate, qualitative statements, possibly for upper bounds for
oscillatory quantities, because we have not specified the precise
measures to use, nor the dependence on (or maximization of the
decay rate over) the initial conditions.

In this work, we show by an explicit calculation for a solvable
pure-decoherence model of two qubits
(two-state systems, spins-$1/2$) interacting with a
bath of bosonic modes, that the measure of entanglement
introduced in [\onlinecite{Wootters}], is indeed suppressed by the
factor $\delta^{\rm (1)} \delta^{\rm (2)}$. We focus on the
two-qubit system, because it is only for this simplest case that
an explicit expression for a measure of entanglement called
concurrence was obtained [\onlinecite{Wootters}]. Our study
expands the recent works [\onlinecite{Eberly,Eberly1}] that considered
similar properties for different models. We are able to derive explicitly
the product of suppression factors result.

For brevity, from now on we will use subscripts or superscripts
$r=1,2$ to label the spins (two-level subsystems),
$H_S^r=\mathcal{A}^r\sigma _z^r$. Each spin interacts with a
bosonic bath of modes $H_B^r=\sum_k\omega _k^rb_k^{r\dagger
}b_k^r$, which has been widely used [\onlinecite{Leggett,
vanKampen, Adiabatic}] as a model of quantum noise (we set $\hbar
=1$). The interaction between the quantum systems and the
environment is taken in the form $H_I^r=\sigma _z^r\sum_k \left(
g_k^{r*}b_k^r+g_k^rb_k^{r\dagger }\right) $. This choice,
corresponding to $[H_B^r,H_I^r]=0$, leads to a solvable model and
has been identified as an appropriate description of pure
decoherence [\onlinecite{Adiabatic}].

We assume that there is no interaction between the
qubits, so that the Hamiltonian of the whole system has the form
$H=\sum_r\left( H_S^r+H_B^r+H_I^r \right)$. The main reason for this assumption is,
of course, to have a solvable model. In addition, we point out
that qubit-qubit interactions, either direct or those induced by the bath modes,
can decrease or {\it increase\/} their entanglement. For the latter reason, we also assumed
that the noise is uncorrelated at the two qubit locations, namely the bath modes
are independent for each qubit (the most natural situation is when
the qubits are spatially separated).

The initial state of the two qubits, described by the density
matrix $\rho_S(0)$, can be entangled. However, we assume
[\onlinecite{Leggett, Adiabatic}] that the qubits are initially
not entangled with the bath modes. The overall initial density
matrix is then
\begin{equation}
\rho \left( 0\right) =\rho _S(0)\otimes \rho _B^1(0)\otimes \rho
_B^2(0) \; .
\end{equation}
The reservoirs are in thermal equilibrium at the
temperature $T$ (with $\beta \equiv 1 / k_{\rm B} T$),
\begin{equation}
\rho _B^r (0)=\prod\limits_k
( 1-e^{-\beta \omega _k^r} )
e^{-\beta \omega
_k^rb_k^{r\dagger }b_k^r}  \;   .
\end{equation}

The total Hamiltonian is time-independent, so the reduced density matrix of the two
qubits at time $t\geq 0$ is
\begin{equation}
\rho _S\left( t\right) ={\rm Tr}_B \left[ U\rho \left( 0\right) U^{\dagger } \right]  \; ,
\label{evol1}
\end{equation}
where the evolution operator factorizes, $U=e^{-iHt}=U_1U_2$. The
trace over the bosonic modes of the two baths, ${\rm Tr}_B$ in
(\ref{evol1}), can then be evaluated exactly by using the
techniques of [\onlinecite{Louisell, Tolkunov}].

It is convenient to write the density operator $\rho_S(t)$ in the
matrix form,
\begin{equation}
\rho _S^{\gamma _1^1 \gamma _1^2,\gamma _2^1 \gamma _2^2}(t)\equiv
\left\langle \gamma _1^1 \gamma _1^2\right| \rho _S(t)\left|
\gamma _2^1 \gamma _2^2\right\rangle \; ,
\end{equation}
where $\gamma _q^r=\pm 1$ has two indexes: $r$ labels the qubit,
while $q$ simply indicates whether it marks row or column matrix
element positions. The values $+1$ and $-1$ correspond to the spin
states $\uparrow$ and $\downarrow$, respectively.

%where we use the standard basis in the system space $\left|
%\uparrow \uparrow \right\rangle ,\left| \uparrow \downarrow
%\right\rangle ,\left| \downarrow \uparrow \right\rangle ,\left|
%\downarrow \downarrow \right\rangle $.

After several straightforward transformations, (\ref{evol1}) is
reduced to
\begin{eqnarray}
&&\rho _S^{ \gamma _1^1 \gamma _1^2, \gamma _2^1 \gamma
_2^2}\left( t\right) \\\nonumber &=& e^{i \mathcal{A}^1\left(
\gamma _2^1- \gamma _1^1\right) t+i\mathcal{A}^2\left( \gamma
_2^2- \gamma _1^2\right) t}T_{\phantom{S}}^{ \gamma _1^1 \gamma
_2^1}T_{\phantom{S}}^{ \gamma _1^2 \gamma_2^2}\rho _S^{ \gamma
_1^1 \gamma _1^2, \gamma _2^1 \gamma _2^2}\left( 0\right) \; ,
\end{eqnarray}
where the coefficients are
\begin{equation}
T^{\gamma _1^r\gamma _2^r}={\rm Tr}_{B_r}\left[ e^{-i\left(
H_B^r+\gamma _1^r \widetilde{H}_I^r\right) t}\rho _B^re^{i\left(
H_B^r+\gamma _2^r\widetilde{H} _I^r\right) t}\right]  \; ,
\end{equation}
here $\widetilde{H}_I^r$ is defined by $H_I^r=\sigma
_z^r\widetilde{H}_I^r$.  Utilizing the identities from
[\onlinecite{Louisell, Tolkunov}], we find an explicit expression
\begin{equation} \label{Tss}
T^{\gamma _1^r\gamma _2^r}=\exp \! \big[ -G_r(t) \left( \gamma
_1^r-\gamma _2^r\right) ^2\big]  \;  ,
\end{equation}
where $G_r(t)$ is the well-studied spectral function [\onlinecite{Palma,
vanKampen}],
\begin{equation}
G_r(t) =2\sum\limits_k\frac{\left| g_k^r\right| ^2}{\left( \omega
_k^r\right) ^2}\sin ^2\frac{\omega _k^rt}2\coth \frac{\beta \omega
_k^r}2 \; .
\end{equation}
A general property of the pure-decoherence models
[\onlinecite{Adiabatic}] is that the diagonal elements of the
density matrix will stay unchanged during the evolution.

Utilizing the new variables $ p_r\equiv e^{2i\mathcal{A}^rt}$ and
$q_r\equiv e^{-4G_r(t) }$ the density matrix can be written explicitly,
\begin{widetext}
\begin{eqnarray}\label{genevol}
\rho_S\left( t\right) = \left(
\begin{array}{cccc}
\rho _S^{\uparrow \uparrow ,\uparrow \uparrow }\left( 0\right)  &
p_2^{*}q_2\rho _S^{\uparrow \uparrow ,\uparrow \downarrow }\left(
0\right) & p_1^{*}q_1\rho _S^{\uparrow \uparrow ,\downarrow
\uparrow }\left( 0\right) & p_1^{*}q_1p_2^{*}q_2\rho _S^{\uparrow
\uparrow ,\downarrow \downarrow
}\left( 0\right)  \\
p_2q_2\rho _S^{\uparrow \downarrow ,\uparrow \uparrow }\left(
0\right)  & \rho _S^{\uparrow \downarrow ,\uparrow \downarrow
}\left( 0\right)  & p_1^{*}q_1p_2q_2\rho _S^{\uparrow \downarrow
,\downarrow \uparrow }\left( 0\right)  & p_1^{*}q_1\rho
_S^{\uparrow \downarrow ,\downarrow \downarrow
}\left( 0\right)  \\
p_1q_1\rho _S^{\downarrow \uparrow ,\uparrow \uparrow }\left(
0\right)  & p_1q_1p_2^{*}q_2\rho _S^{\downarrow \uparrow ,\uparrow
\downarrow }\left( 0\right)  & \rho _S^{\downarrow \uparrow
,\downarrow \uparrow }\left( 0\right)  & p_2^{*}q_2\rho
_S^{\downarrow \uparrow ,\downarrow \downarrow
}\left( 0\right)  \\
p_1q_1p_2q_2\rho _S^{\downarrow \downarrow ,\uparrow \uparrow
}\left( 0\right)  & p_1q_1\rho _S^{\downarrow \downarrow ,\uparrow
\downarrow }\left( 0\right)  & p_2q_2\rho _S^{\downarrow
\downarrow ,\downarrow \uparrow }\left( 0\right)  & \rho
_S^{\downarrow \downarrow ,\downarrow \downarrow }\left( 0\right)
\end{array}
\right)  \; .
\end{eqnarray}
\end{widetext}

\begin{figure}[tbp]
\includegraphics[width=9cm, height=9cm]{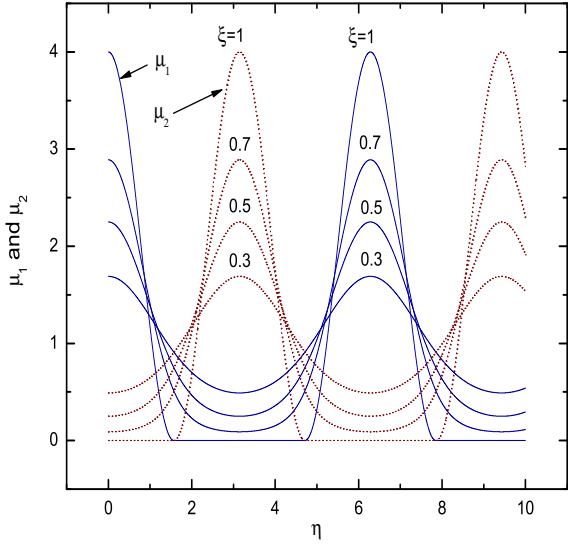}
\caption{Eigenvalues $\mu_1$ and $\mu_2$ as functions of $\eta$,
for several values of $\xi$, with the prefactor $|\alpha|^2 /
(1 + |\alpha|^2  )^2$ suppressed. } \label{graph1}
\end{figure}

To analyze the effect of decoherence on the entangled qubit states
we use a measure of entanglement. The entanglement of formation
[\onlinecite{Bennett1}] was historically the first widely accepted
measure of entanglement. For a mixed state $\rho_S$, the
evaluation of this measure is related to minimization over all the
possible pure-state decompositions of $\rho_S$, and even for a
two-qubit system getting analytical results for this measure is a
complicated problem. The {\it concurrence\/}
[{\onlinecite{Wootters}] is a quantity monotonically related to
the entanglement of formation, hence it may be used as a
convenient substitute for it.
Given a pure or mixed state, $\rho_S$, of
two qubits, we define the spin-flipped state
\begin{gather}
\widetilde{\rho}_S=\left( \sigma _y\otimes \sigma _y\right) \rho_S
^{*}\left( \sigma _y\otimes \sigma _y\right) \; ,
\end{gather}
and the Hermitian matrix
$R(\rho_S)=\sqrt{\sqrt{\rho_S}\,\widetilde{\rho}_S\sqrt{\rho_S}}$ with
eigenvalues $\lambda_{i=1,2,3,4}\thinspace$. Then the concurrence
[\onlinecite{Wootters}] is given by
\begin{gather}\label{conc}
C\left( \rho _S\left( t\right) \right) =\max \big\{
0,\;2\max\limits_i \lambda _i - \sum\limits_{j=1}^4\lambda
_j\big\} \; .
\end{gather}
Since we know $\rho_S(t)$ explicitly (\ref{genevol}), the
evaluation of (\ref{conc}) is reduced to finding the eigenvalues of
a $4\times4$ matrix.

For illustration, we considered the system of two qubits in a pure
state which at time \hbox{$t=0$} is $\left| \psi \right\rangle
=\left( \left| \uparrow \downarrow \right\rangle +\alpha \left|
\downarrow \uparrow \right\rangle \right) /\sqrt{1+|\alpha| ^2}$.
Here the (complex) parameter $\alpha $ characterizes the degree of
entanglement. Under the influence of the quantum noise the system
evolves from the state $\rho _S\left( 0\right) =\left| \psi
\right\rangle \left\langle \psi \right|$ to the mixed state
\begin{gather}
\rho _S\left( t\right) =\frac 1{1+\left|\alpha\right| ^2}\left(
\begin{array}{cccc}
0 & 0 & 0 & 0 \\
0 & 1 & p_1^{*}q_1p_2q_2\alpha^* & 0 \\
0 & p_1q_1p_2^{*}q_2\alpha & | \alpha |^2 & 0 \\
0 & 0 & 0 & 0
\end{array}
\right) \; .
\end{gather}

To evaluate the measure of entanglement at times $t>0$, we have to
find the eigenvalues of the matrix $R$, which can be obtained from
the eigenvalues of the product $\rho _S\left( t\right) \widetilde{\rho }_S\left(
t\right)$. The latter eigenvalues are
\begin{equation}
\mu _{1,2}=\frac{|\alpha| ^2}{\left( 1+| \alpha |^2\right) ^2}
\bigg(1+ \xi ^2\cos 2\eta \pm 2\xi \cos \eta \sqrt{1-\xi ^2\sin
^2\eta }\, \bigg)  \; ,
\end{equation}
and $\mu _{3,4}=0$. Here $\xi \equiv e^{-4\left[ G_1(t) +G_2(t) \right] }$
and $\eta
\equiv 2\left( \mathcal{A}^2-\mathcal{A}^1\right) t$. Then the eigenvalues $\lambda_i = \sqrt{\mu_i}\,$, and as
a result the concurrence takes the form,
\begin{gather}
C\left( \rho _S\left( t\right) \right) =\;\left| \sqrt{\mu
_1}-\sqrt{ \mu _2}\right| \; .
\end{gather}
The eigenvalues $\mu_{1,2}$ are shown in Fig.~1.
For example, for a simple case of identical qubits, $\mathcal{A}^2=\mathcal{A}^1$ ,
we have $\eta =0 $ and $\lambda _{1,2}=| \alpha |( \xi \pm1)
/(1+|\alpha |^2)$. As a result the concurrence is
\begin{gather}
C_{\eta=0} =\frac{2 | \alpha |}{1+| \alpha |^2}\;e^{-4\left[ G_1(t)
+G_2(t)\right] } \; .
\end{gather}
This establishes the product of the suppression factors property
alluded to in the introduction, because it is known
[\onlinecite{Adiabatic}] that each of the exponential factors
$e^{-4G_{1,2}(t)}$ measures the decay of the off-diagonal matrix
elements when each qubit is isolated from the other, but exposed
to its own bath. When the qubits are not identical, one can prove
that for any $t\geq 0$,
\begin{gather}\label{ineq}
C_{\eta}(t) \le C_{\eta=0}(t)\; ,
\end{gather}
so that the product of the factors property applies as an upper
bound. The recent Markovian-approximation results
[\onlinecite{Eberly1,Eberly2}], appropriate for large times, have
yielded an interesting observation that for some initial
conditions the concurrence, unlike coherence, can drop to zero in
finite time [\onlinecite{F3,F2,F1}]. We have not explored this
property within the pure-decoherence scheme considered here.

In summary, we connected two important issues in the studies of
entanglement and decoherence, namely, for a solvable
pure-decoherence model, we confirmed that the decay of
entanglement is approximately governed by the product of the
suppression factors describing decoherence of the subsystems,
provided that they are subject to uncorrelated sources of noise.
Our results also suggest avenues for future work. Specifically,
for multiqubit systems, one might speculate that similar arguments
could apply ``by induction.'' However, understanding of
entanglement is far from intuitive, especially when one considers
more than two two-state systems. Therefore, for any definitive
progress, one has first to develop appropriate quantitative
measures of entanglement, and try to quantify entanglement and
decoherence in a unified way.

This research was supported by the National Security
Agency and Advanced Research and Development Activity under Army
Research Office contract DAAD-19-02-1-0035, and by the National
Science Foundation, grant DMR-0121146.

\end{document}